\documentclass[twocolumn,showpacs]{revtex4}

\usepackage{graphicx}
\usepackage{amsmath}
\usepackage{amssymb}

\begin{document}

%\preprint{HEP/123-qed}

\title{Low temperature superfluid stiffness of $d$-wave superconductor
in a magnetic field}

\author{S.G.~Sharapov$^{1,*}$}
%\thanks{On leave of absence from Bogolyubov
%         Institute for Theoretical Physics, Kiev, Ukraine}%
%\email{Sergei.Sharapov@unine.ch}
\author{V.P.~Gusynin$^2$}
%\email{vgusynin@bitp.kiev.ua}
\author{H.~Beck$^1$}%
%\email{Hans.Beck@unine.ch}
\affiliation{
        $^1$Institut de Physique,
        Universit\'e de Neuch\^atel, 2000 Neuch\^atel, Switzerland\\
        $^2$Bogolyubov Institute for Theoretical Physics,
        Metrologicheskaya Str. 14-b, Kiev, 03143, Ukraine}

\date{\today }

\begin{abstract}
The temperature and field dependence of the superfluid density
$\rho_s$ in the vortex state of a $d$-wave superconductor are
calculated using a microscopic model in the quasiclassical
approximation. We show that at temperatures below $T^{\ast}
\varpropto \sqrt{H}$, the linear $T$ dependence of $\rho_s$
crosses over to a $T^2$ dependence differently from the behavior
of the effective penetration depth, $\lambda_{\rm eff}^{-2}(T)$.
We point out that the expected dependences could be probed by a
mutual-inductance technique experiment.
\end{abstract}


% 74.25.Nf Response to electromagnetic fields (nuclear magnetic resonance,
%           surface impedance, etc.)
% 74.72.-h  High-Tc compounds
% 74.25.Jb      Electronic structure


\maketitle

{\it Introduction.} While the mechanism of superconductivity and
unusual nature of the normal state in high-temperature
superconductors (HTSC) are not yet understood, there is a
consensus that the zero field superconducting state has a $d$-wave
superconducting energy gap, with nodes along the diagonals of the
Brillouin zone \cite{Tsuei:2000:RMP}. The presence of the nodes
results in a density of low energy quasiparticle excitations large
compared with conventional $s$-wave superconductors even at the
temperatures much smaller than the transition temperature, $T \ll
T_c$. Although these excitations are reasonably well described by
Landau quasiparticles their presence brings in a qualitatively new
quasiparticle phenomenology not encountered in conventional
superconductors. Among many aspects of this new physics a major
role is played by these low energy excitations in the mixed (or
vortex) state
\cite{Krishana:1997:SCI,Lee:1997:SCI,Vekhter:2001:PRB,Vafek:2001:PRB}.
Since all HTSCs are extreme type-II superconductors a huge mixed
phase extends from the lower critical field, $H_{c1} \sim 10 - 100
\, \mbox{Gauss}$ to the upper critical field, $H_{c2} \sim 100 \,
\mbox{T}$. An important property of the vortex state was pointed
out by Volovik \cite{Volovik:1993:JETPL}, who predicted, that in
contrast to conventional superconductors, in $d$-wave systems the
density of states (DOS) is dominated by contribution from excited
quasiparticle states rather than the bound states associated with
vortex cores. It was shown that in an applied magnetic field
$H$ the extended quasiparticles DOS, $N(\omega =0, H) \sim
\sqrt{H}$ rather than $\sim H$ as in the conventional case. This
result was confirmed by specific heat measurements on high quality
single crystals \cite{Moler:1994:PRL,Wang:2001:PRB}. Subsequently
the semiclassical treatment of \cite{Volovik:1993:JETPL} was
incorporated into the Green's functions formalism extended to include the
effects of impurity scattering \cite{Kubert:1998:SSC}. Accounting
for the impurity scattering which violates a simple $\sqrt{H}$
dependence has improved the agreement between the theory and
measurements of the electronic specific heat (see Refs. in
\cite{Vekhter:2001:PRB}).

A similar to the Volovik effect, the weak ($H < H_{c1}$) field
response was studied by Yip and Sauls \cite{Yip:1992:PRL}. They
predicted a direction dependent {\it nonlinear} Meissner effect
also associated with quasiclassical shift of the excitation
spectrum due to the superflow created by the screening currents.
Although initially an experimental evidence for such an effect was
reported, the subsequent experiments \cite{Carrington:1999:PRB}
did not confirm this effect.

Another interesting result was obtained by the $\mu$SR measurements of
temperature and field dependences of the effective in-plane
penetration depth, $\lambda_{\rm eff}$ in single crystals
YBa$_2$Cu$_3$O$_{6.95}$ by Sonier {\em et al.}
\cite{Sonier:1999:PRL}. At high magnetic fields,
they observed a flattening of $\lambda_{\rm eff}^{-2}$ (defined in
these experiments as the width of magnetic field distribution) at
low temperatures in contrast to the linear $T$ behavior expected
in a clean $d$-wave superconductor. If one assumes $\lambda_{\rm
eff}^{-2}$ is proportional to the superfluid density, then such a
flattening could, in principle, indicate an opening of a secondary
gap at the nodes of a $d$-wave superconducting gap as was already
suggested after the measurements of thermal conductivity
\cite{Krishana:1997:SCI}.

It was argued by Amin {\em et al.} \cite{Amin:2000:PRL} that the
simple relation $\lambda_{\rm eff}^{-2} \varpropto \rho_{s}$ is
not valid for the penetration depth extracted in $\mu$SR
experiments at finite fields, so that using the proper definition
\cite{Amin:1998:PRB} of $\lambda_{\rm eff}$, which corresponds to
the $\mu$SR measured penetration depth, the observed behavior of
$\lambda_{\rm eff}^{-2}$ can be explained by a nonlocal London
model for a $d$-wave superconductor. The behavior of the
superfluid stiffness, defined as a second derivative of the free
energy of the system with respect to a vector potential,
in this case cannot be addressed by the $\mu$SR.

Nevertheless the superfluid density in itself could also be
extracted, for example, from the measurements of the low frequency complex
sheet conductance as done in two-coil mutual-inductance
technique measurements on thin films
\cite{Jeanneret:1989:APL,Calame:2001:PRL}. The
superfluid density (stiffness) in this case is found from the
inductive part $\sigma_2(\omega)$ of the conductivity
\cite{Tinkham.book}
\begin{equation}\label{basic.def}
\frac{\rho_s(T)}{m} \equiv \frac{c^2}{4 \pi e^2 \lambda_{L}^2(T)} =
\frac{1}{e^2} \lim_{\omega \to 0} \omega \sigma_2(\omega,T),
\end{equation}
where $e$ is the electron charge, $m$ is its mass (or, more
generally, a mass of the charge carrier) and $c$ is the velocity
of light. It is very important to make a distinction between the
London penetration depth $\lambda_{L}$ appearing in
(\ref{basic.def}) and $\lambda_{\rm eff}$ deduced from the $\mu$SR
measurements \cite{Amin:1998:PRB,Wang:1999:SSC}.

Thus in this work we investigate the influence of the magnetic
field directly on the superfluid stiffness in HTSC using the
quasiclassical approximation at $H_{c1} < H \ll H_{c2}$.
Our result is that at low temperatures
there is also a flattening of $\rho_s$ which is not related to
an opening of the gap and is just the result of a Doppler
shift of the quasiparticle energies in the vortex state.

Although a quantum mechanical treatment is certainly needed for
a consistent treatment of the vortex state \cite{Vafek:2001:PRB},
the semiclassical approximation always provides a good starting
point when a new physical property of the system is involved.
Furthermore, as argued in \cite{Vekhter:2001:PRB} for the
parameter range relevant to the study of HTSC, this approximation
reproduces the energy spectrum of the near-nodal quasiparticles in
a vortex state to a high degree of accuracy allowing also to
include the effect of impurity scattering into the analysis. In
particular, the thermal conductivity was studied using the
semiclassical approximation
\cite{Kubert:1998:PRL,Franz:1999:PRL,Vekhter:1999:PRL}.

{\em Superfluid stiffness in the presence of impurities.}
Superfluid stiffness is given by \cite{Durst:2000:PRB}:
\begin{equation}\label{stiffness.def}
\frac{\rho_{s}^{\alpha \beta} (T,H)}{m} =
\frac{\rho_{s 0}^{\alpha \beta}}{m} - \frac{\rho_{n}^{\alpha \beta} (T,H)}{m}.
\end{equation}
The normal fluid density, $\rho_n$ in (\ref{stiffness.def}),
calculated within the ``bubble approximation'' with dressed
fermion propagators (i.e., with self-energy $\Sigma$ due to the
scattering on impurities included) but neglecting the vertex and
Fermi liquid corrections, is
\begin{equation}\label{rho.n}
\begin{split}
& \frac{\rho_{n}^{\alpha \beta}}{m} =
\int \frac{d^2 k}{(2 \pi)^2} \int_{-\infty}^{\infty} d
\omega  \tanh \frac{\omega}{2 T}
\frac{ v_{F \alpha} v_{F \beta}}{4 \pi i} \\
& \times \mbox{tr}[
G_{A}(\omega, \mathbf{k}) G_{A}(\omega,\mathbf{k})) -
G_{R}(\omega, \mathbf{k}) G_{R}(\omega, \mathbf{k})] .
\end{split}
\end{equation}
As shown in  \cite{Durst:2000:PRB} the vertex corrections can be neglected
if the impurity scattering potential is isotropic in $\mathbf{k}$-space.
Likewise the Fermi liquid corrections can be taken into account
along the lines of Ref.\cite{Durst:2000:PRB}.
In (\ref{rho.n})
$G_{R,A}(\omega,\mathbf{k})$ are the retarded and advanced Green's
functions
\begin{equation}
\label{Green.dirty}
G_{R,A} (\omega, \mathbf{k})  =
\frac{(\omega \pm i \Gamma) \hat{I} +
\tau_3 \xi(\mathbf{k}) - \tau_1 \Delta(\mathbf{k})}
{(\omega \pm i \Gamma)^2 - \xi^2(\mathbf{k}) - \Delta^2 (\mathbf{k})}
\end{equation}
with the dispersion law $\xi(\mathbf{k})$, the $d$-wave
superconducting gap $\Delta(\mathbf{k})$, and the Fermi velocity
is $\mathbf{v}_{F} = \partial \xi (\mathbf{k})/\partial
\mathbf{k}|_{\mathbf{k} = \mathbf{k}_F}$. The width $\Gamma =
-\mbox{Im}\Sigma(\omega)$ in (\ref{Green.dirty}) is the scattering
rate due to impurities and other sources, e.g. disordered vortex
lattice (see below). Following \cite{Durst:2000:PRB} we assume
that around the nodes $\Gamma$ is frequency and momentum
independent, so that the presence of impurities is modeled by a
widening of $\delta$-like quasiparticle peaks by a Lorentzian
\begin{equation}
\begin{split}
& A(\omega, \mathbf{k}) = \frac{1}{2\pi i}
[G_{A} (\omega- i 0, \mathbf{k}) -
G_{R} (\omega+ i 0, \mathbf{k})] \\
& = \frac{\Gamma}{2 \pi E} \left[ \frac{E
+ \tau_3 \xi - \tau_1 \Delta}{(\omega - E)^2 + \Gamma^2} + \frac{E
- \tau_3 \xi + \tau_1 \Delta}{(\omega + E)^2 + \Gamma^2} \right].
\end{split}
\end{equation}
with a constant width $\Gamma$ (see, however,
\cite{Loktev:2001:FNT} where this assumption was not confirmed for
dopants considered as impurity centers). Here $E(\mathbf{k}) =
\sqrt{\xi^2 (\mathbf{k}) + \Delta^2(\mathbf{k})}$ is the
quasiparticle dispersion law.

Employing for the low temperature regime, $T \ll T_c$ the nodal
approximation \cite{Durst:2000:PRB} we obtain that
\begin{equation}
\label{rho.final.dirty}
\frac{\rho_{n}^{\alpha \beta}(T)}{m} =
\frac{v_{F} \delta_{\alpha \beta}}{\pi v_{\Delta}} J,
\end{equation}
where $\mathbf{v}_{\Delta} =
\partial \Delta (\mathbf{k})/\partial \mathbf{k}|_{\mathbf{k} = \mathbf{k}_F}$
is the gap velocity and
\begin{equation}
\label{J.H=0}
\begin{split}
J=& \frac{1}{\pi}\int\limits_{0}^\infty
d\omega\tanh\frac{\omega}{2T} \\
\times & \left[ \arctan\frac{\Gamma^2-\omega^2}{2\omega\Gamma} -
\arctan \frac{\Gamma^2-\omega^2+p_0^2}{2\omega\Gamma}
\right. \\
&-  \left.\frac{p_0\Gamma}{(p_0+\omega)^2+\Gamma^2}+
\frac{p_0\Gamma}{(p_0-\omega)^2+\Gamma^2}\right]
\end{split}
\end{equation}
with the cutoff energy $p_0$ (as estimated in
\cite{Vekhter:2001:PRB} $p_0 \sim 1500 \mbox{K}$).
Taking the no
impurities limit, $\Gamma \to 0$ one can recover from
(\ref{rho.final.dirty}), (\ref{J.H=0}) the known linear $T$
dependence \cite{Durst:2000:PRB}
\begin{equation}
\label{rho.clean}
\frac{\rho_{n}(T)}{m} = \frac{2 \ln 2}{\pi}
\frac{v_{F} }{v_{\Delta}} T,
\end{equation}
where we set $\hbar = k_B = 1$.
It is essential, however, to take into account
the presence of impurities which modifies the low temperature $T$
dependence of $\rho_n$ from linear to $\sim T^2$
\cite{Hirschfeld:1993:PRB,Xiang:1998:IJMPB}. Indeed in the
limit $\Gamma \gg T$ we get from  (\ref{rho.final.dirty}), (\ref{J.H=0}) that
\begin{equation}\label{rho.dirty.final2}
\frac{\rho_{n}(T)}{m} =
\frac{v_{F}}{\pi v_{\Delta}}
\left[\frac{2 \Gamma}{\pi} \ln \frac{p_0}{\Gamma} +
\frac{\pi}{3} \frac{T^2}{\Gamma} - O\left(\frac{T^4}{\Gamma^3}\right)
\right].
\end{equation}
For clean YBCO monocrystals the estimated \cite{Ando:2000:PRB} value of
the scattering due to impurities
$\Gamma_0 \sim 1-2 \mbox{K}$, but in thin films the quadratic temperature
dependence of $\rho_s(T)$ is observed over a wide temperature range
\cite{Calame:2001:PRL} implying that the value of $\Gamma_0$ could be
bigger, so that for our estimates we use $\Gamma_0 = 10 \mbox{K}$.
However, as we will see below the obtained dependences on $H$ are in fact
even stronger for smaller values of $\Gamma_0$.

{\em Doppler shift.}
In the semiclassical approach to the vortex state the presence of a
superflow is accounted for by introducing the Doppler shift into the
energy $\omega \to \omega - \epsilon(\mathbf{k}, \mathbf{r})$
\cite{Volovik:1993:JETPL,Kubert:1998:SSC}, where
$\epsilon(\mathbf{k}, \mathbf{r}) = \mathbf{v}_s (\mathbf{r})
\cdot \mathbf{k}$, and $\mathbf{v}_s (\mathbf{r})$ is the supervelocity
field at a position $\mathbf{r}$ created by all vortices. There are several
ways to treat the problem depending on the assumptions
we made about the vortex lattice.

1. {\em Completely disordered vortex state.}
Following \cite{Franz:1999:PRL} we assume that the vortex
lattice is {\em disordered}, which is not unreasonable for thin films
\cite{Calame:2001:PRL}. Then on the length scales large compared with
the intervortex distance $a_v = \sqrt{\Phi_0/\pi B}$ (here $\Phi_0 = h c/2e$
is the flux quantum and $B$ is the internal field),
propagation of quasiparticles is described by the
Green's function averaged over the vortex positions $\{\mathbf{R}_i\}$.
This averaging can be done using the probability density
\begin{equation}
\label{P.Franz}
\mathcal{P}(\eta) = \langle
\delta(\eta - \mathbf{v}_s(\mathbf{r}) \cdot \mathbf{k})
\rangle_{\{\mathbf{R}_i\}},
\end{equation}
so that the averaged Green's function
\begin{equation}
\label{Green.P} \langle G(\omega, \mathbf{k}) \rangle =
\int_{-\infty}^{\infty} d \eta \mathcal{P}(\eta) G^{0}(\omega -
\eta, \mathbf{k}),
\end{equation}
where $G^{0}(\omega, \mathbf{k}) = G(\omega, \mathbf{k})$ from Eq.~(\ref{Green.dirty})
with $\Gamma = 0$. For typical accessible fields $H_{c1} < H \ll H_{c2}$
the magnetization due to the vortex lattice is small, and the internal
field $B$ can be replaced by the applied field, $H$ directed along
the $c$-axis.

Assuming that the vortex positions are random
and uncorrelated, one would get \cite{Yu:1995:PRL}
a Gaussian distribution
$\mathcal{P}(\eta) = (2 \pi E_H^2)^{-1/2}
e^{-\eta^2/2 E_H^2}$ with
\begin{equation}
\label{E_H}
E_{H}^2 = k_{\alpha} k_{\beta}
\langle v_{s}^{\alpha}(\mathbf{r}) v_{s}^{\beta}(\mathbf{r})\rangle
\approx \left( \frac{\hbar v_F}{2 a_v} \right)^2 =
\frac{\pi}{4} \hbar^2 v_F^2 \frac{H}{\Phi_0},
\end{equation}
where we used the expression
$|\mathbf{v}_s(\mathbf{r})| = \hbar/2 m r$ (the Planck's constant $\hbar$
is restored in these expressions) for a supervelocity field created by a
single vortex and took $r = a_v$. In what follows we will use an estimate
$E_{H}[\mbox{K}] \sim 30 \sqrt{H [\mbox{T}]}$
obtained for the typical values of
$v_{F} \approx (1.5 - 2.5)\times 10^7 \mbox{cm/s} $ in YBCO.
To make an analytical calculations possible we follow
Ref.\cite{Franz:1999:PRL} and
replace the Gaussian probability distribution
by a Lorentzian
$\mathcal{P} (\eta) = \pi^{-1} E_H/(\eta^2 + E_H^2)$, so
that
\begin{equation}
\label{Green.H}
\langle G_{R}(\omega, \mathbf{k}) \rangle =
G_{R}^{0}(\omega+ i E_H, \mathbf{k}).
\end{equation}
Thus within employed approximation the effect of quasiparticle
scattering from the supervelocity field $\mathbf{v}_s(\mathbf{r})$
created by a completely disordered vortex lattice also results in
a widening of the quasiparticle peaks with a constant width $E_H$.
Although this property may not be valid for other forms of the
distribution functions, we find it useful for a physical insight to the
problem.

We note that the used above Lorentzian distribution has another property
that the average of the product of the Green's functions is equal
to the product of averaged Green's functions, so that the individual Green's
function can be averaged over the vortex positions.
For any other type of the distribution function this property is not valid and
averaging should be done for the product of the Green's functions itself (see below).

Eq.~(\ref{Green.H}) may be used to calculate the field induced
DOS, $N(\omega, H) = 1/(8 \pi^2) \int d^2 k \mbox{tr} A(\omega,
\mathbf{k})$, so that $N(\omega =0, H) \varpropto E_H \ln
p_0^2/E_H^2 \sim \sqrt{H} \ln H_0/H$ with $H_0 \sim 2500
\mbox{T}$. This is just DOS from \cite{Durst:2000:PRB} with
$\Gamma$ replaced by $E_H$. Since $p_0 \gg E_H$ this result agrees
with the dependence $N (\omega =0, H) \sim \sqrt{H}$ obtained by
Volovik \cite{Volovik:1993:JETPL}. Furthermore, as we will see
below, the extra $\ln$ factor is caused by the assumption we made
about the completely disordered vortex state. Note also, that the
first temperature independent term of Eq.~(\ref{rho.dirty.final2})
may be identified as a DOS contribution to $\rho_n$.

Assuming also a Matthiessen type rule for the impurity and vortex
contributions to the lifetime \cite{Yu:1995:PRL}, we finally obtain
that the field dependence of the normal fluid density is described by
Eqs.~(\ref{rho.final.dirty}), (\ref{J.H=0}) with $\Gamma = \Gamma_0 + E_H$.
and its asymptotics for $T \gg \Gamma$ and $T \ll \Gamma$ are given
by Eqs.~(\ref{rho.clean}) and (\ref{rho.dirty.final2}), respectively.
\begin{figure}
\centering{
\includegraphics[width=8.5cm]{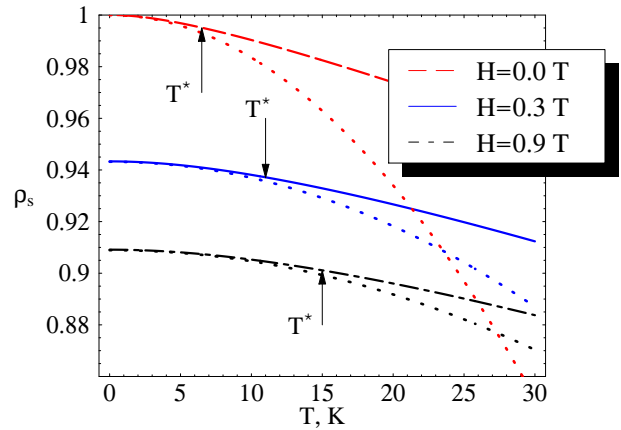}}
\caption{The temperature dependences of $\rho_{s}(T,H)/\rho_s(0,0)$
calculated from Eqs.~(\ref{stiffness.def}), (\ref{rho.final.dirty}),
(\ref{J.H=0}) and $\Gamma = \Gamma_0 + E_H$.
The dotted lines are calculated using the $T^2$ asymptotic
(\ref{rho.dirty.final2}).}
\label{fig:1}
\end{figure}
These results are presented in Fig.~\ref{fig:1} (we
used for computation some typical values of the parameters
\cite{Chiao:2000:PRB}, e.g. $v_{F}/v_{\Delta} = 15$),
where one can clearly see a crossover from $T^2$
dependence of $\rho_s (T)$ to the linear $T$ dependence at
$T^{\ast} \sim \Gamma_0 + E_H$. For a clean system ($\Gamma_0 =0$)
and a single vortex averaging this crossover of $\rho_s(T)$ was
also obtained in \cite{Vekhter:1999:PRB}.
Thus this behavior of $\rho_s(T)$
indeed differs from the effective penetration depth
$\lambda_{\rm eff}^{-2} (T)$  \cite{Amin:2000:PRL}
which also crosses over from the linear $T$ dependence, but to a
$T^3$ dependence at (for a clean system)
$T^{\ast} \varpropto \sqrt{H}$.
Note also that since the nodal approximation was used to arrive at
Eqs.~(\ref{rho.final.dirty}), (\ref{J.H=0}) one cannot approach
the region  $\rho_{s}(T,H)/\rho_s(0,0) \ll 1$, where the dependence
of the superconducting gap on temperature should be taken into account.

\begin{figure}
\centering{
\includegraphics[width=7.5cm]{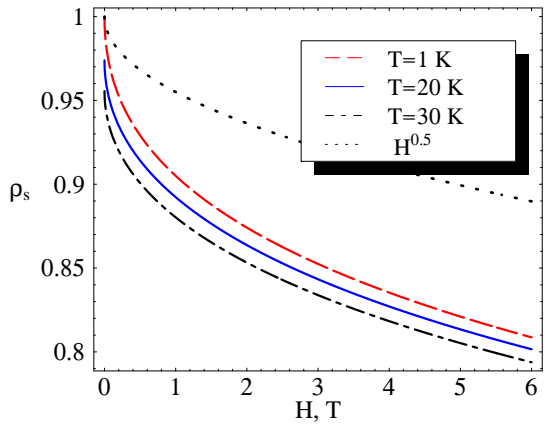}}
\caption{The dependence $\rho_{s}(T,H)/\rho_s(0,0)$ on the applied field
calculated
from Eqs.~(\ref{stiffness.def}), (\ref{rho.final.dirty}),
(\ref{J.H=0}) and $\Gamma = \Gamma_0 + E_H$.
The dotted line is obtained using the first term of
Eq.~(\ref{vortex.glass.final}),
i.e. for $\rho_n = v_{F}/(\pi v_{\Delta}) E_H$. }
\label{fig:2}
\end{figure}

In Fig.~\ref{fig:2} we show the dependence of $\rho_s$ on the applied
field $H$. It appears to be not very different from $\sim \sqrt{H}$
dependence obtained in \cite{Ioffe:2001} for the periodic vortex lattice at
$T = 0$. The reason for this coincidence becomes more clear after we
consider a vortex liquid case.

2. {\em Vortex liquid.} To consider a more ordered vortex liquid
state we adopt a method used in \cite{Vekhter:2001:PRB}. In this
case one introduces a Doppler shift $\epsilon_n (\mathbf{r}) =
\mathbf{v}_s (\mathbf{r}) \cdot \mathbf{k}_n$ at the node
$\mathbf{k}_n$ ($\mathbf{k}_1 = - \mathbf{k}_3$ and $\mathbf{k}_2
= - \mathbf{k}_4$ for a $d$-wave superconductor, so that
$\epsilon_1 = - \epsilon_3$ and $\epsilon_2 = - \epsilon_4$) to
approximate the Doppler shift for the entire nodal region. Then if
we know how to express a physical quantity $F$ in terms of  the
Green function $G(\omega, \mathbf{k})$ we can compute its local
value $F(\mathbf{r})$ with the local
 ``Doppler shifted'' Green's function,
$G(\omega, \mathbf{k}; \epsilon(\mathbf{k}, \mathbf{r})) =
G(\omega - \epsilon_{n} (\mathbf{r}), \mathbf{k})$.
We then approximate the field-dependent
measured value $F(H)$ by the spatial average which depends on the magnetic
field $H$ through $\epsilon_n$ as \cite{Kubert:1998:SSC}
$A^{-1} \int d^2 \mathbf{r}
F(\epsilon_1 (\mathbf{r}), \epsilon_2 (\mathbf{r})) =
\int_{-\infty}^{\infty} d \epsilon_1  d \epsilon_2
F(\epsilon_1, \epsilon_2) \mathcal{L} (\epsilon_1, \epsilon_2)$,
where the first integral is taken over the part of a unit cell of the vortex
lattice (with the area $A$) in real space where the Doppler shift
is much smaller than the gap maximum and in the second integral
we used the distribution function
\begin{equation}
\mathcal{L} (\epsilon_1, \epsilon_2) = A^{-1} \int d^2 \mathbf{r}
\delta(\epsilon_1 - \mathbf{v}_s (\mathbf{r}) \mathbf{k}_1)
\delta(\epsilon_2 - \mathbf{v}_s (\mathbf{r}) \mathbf{k}_2),
\end{equation}
where $\mathbf{k}_1$ and  $\mathbf{k}_2$ label two nearest nodes.
If $\mathcal{L} (\epsilon_1, \epsilon_2)$ depends on a single
variable $\epsilon_1^2 + \epsilon_2^2$
then a more simple distribution function $\mathcal{P} (\epsilon) =
\int_{-\infty}^{\infty} d \epsilon_1 \mathcal{L} (\epsilon, \epsilon_1)$
may be used. Among several candidates for $\mathcal{P} (\epsilon)$
suggested in \cite{Vekhter:2001:PRB}
we choose $\mathcal{P}(\epsilon) = 1/2 E_H^2(\epsilon^2 + E_H^2)^{-3/2}$
which is the most convenient for analytical calculations.
Using this function one can obtain the following field induced
DOS: $N(\omega = 0, H) \varpropto \sqrt{H}$ in the strong field,
$\Gamma_0 \ll |\epsilon_n|$ limit and
$N(\omega = 0, H) \varpropto H \ln H_0/H$ in the impurity
dominated, $\Gamma_0 \gg |\epsilon_n|$ regime. This shows that
this distribution function which allows to reproduce the original
$\sim \sqrt{H}$ Volovik's result is indeed better for the description
of a more ordered vortex glass or even lattice state.

Hence the local quantity $J$ from Eq.~(\ref{J.H=0})
has to be replaced by its local value $J (\epsilon_n)$ and
\begin{equation}
\label{rho.final.lattice}
\frac{\rho_n}{m}
 = \frac{v_F }{\pi v_{\Delta}}
\int_{-\infty}^{\infty} d \epsilon J(\epsilon)
\mathcal{P} (\epsilon).
\end{equation}
One can obtain that the impurity dominated limit
($T \ll |\epsilon_n| \ll \Gamma_0$) is described by Eq.~(\ref{rho.dirty.final2})
with $\Gamma = \Gamma_0$, while in the field dominated,
$T \ll  \Gamma_0 \ll |\epsilon_n|$ regime
\begin{equation}
\label{vortex.glass.final}
\frac{\rho_n}{m} =
\frac{v_{F}}{\pi v_{\Delta}} \left[
E_H + \frac{2 \Gamma_0}{\pi} \ln \frac{2 p_0}{E_H} + \frac{\pi T^2}{3}
\frac{\pi }{2 E_H}
\right].
\end{equation}
Thus irrespectively of the vortex structure we assumed the last
$\sim T^2$ term of (\ref{vortex.glass.final}) differs only by a
factor $\pi/2$ from Eq.~(\ref{rho.dirty.final2}) taken with $\Gamma
= \Gamma_0 + E_H \sim E_H$ for the field dominated regime, $E_H
\gg \Gamma_0$. It also agrees with \cite{Vekhter:1999:PRB}, where,
however, there is no  second term related to impurities. The first
term of  (\ref{vortex.glass.final}) is already $\sim \sqrt{H}$,
not $\sim \sqrt{H} \ln H_0/H$, exactly as one would expect from
the DOS calculation and obtains for a single vortex averaging
\cite{Vekhter:1999:PRB} and the vortex lattice as $T = 0$
\cite{Ioffe:2001}.

{\em To conclude} we have shown using the quasiclassical approximation and
the simplest form for the impurity scattering that the
temperature $T^{\ast} \sim \Gamma_0$ of the crossover from the linear to
the $T^2$ dependence of the superfluid stiffness $\rho_s(T,H)$
increases in the presence of magnetic field to the temperature
$T^{\ast} \sim \Gamma_0 + E_H$. The dependence of $\rho_s(T)$ is
different from the effective penetration depth, $\lambda_{\rm eff}^{-2}(T)$.
The experimental measurements of $\rho_s(T,H)$
are necessary to check our predictions against them.

{\em Acknowledgments}.
We gratefully acknowledge V.M.~Loktev and P.~Martinoli for
numerous fruitful discussions. S.G.Sh would also like to thank
C.~Panagopoulos for bringing Ref.~\cite{Sonier:1999:PRL}
to his attention. This work was supported by the
research project 20-65045.01 and by the SCOPES-project
7UKPJ062150.00/1 of the Swiss National Science Foundation.

\end{document}